\documentclass[12pt]{article}%
\usepackage{epsfig}
\usepackage{amsmath}
\usepackage{amsfonts}
\usepackage{amssymb}
\usepackage{graphicx}%
\providecommand{\U}[1]{\protect\rule{.1in}{.1in}}
\begin{document}

\begin{center}
{\Large \textbf{String Scattering Amplitudes}}

{\Large \textbf{in High Energy Limits}}
\end{center}

\bigskip\bigskip


\begin{raggedright}
{\it Yi Yang and Jenchi Lee \index{Yang, Y.}\\
Department of Electrophysicss\\
National Chiao Tung University\\
1001 University Street\\
Hsinchu, Taiwan}
\bigskip\bigskip
\end{raggedright}

\section{Introduction}

Quantum Field Theory (QFT) is a powerful theory in modern physics. Based on
QFT, standard model of particle physics successfully describes our microcosmic
world. Most of the predictions by standard model have been observed in many
experiments under rather precise level. However, the key technical procedure
in QFT, i.e. renormalization, looks complicated and cumbersome. More
seriously, the renormalization procedure does not work for gravity, which
means\ that it is impossible to construct a consistent quantum gravity theory
by using the conventional QFT. Many efforts have been made to understand the
physical meaning of renormalization, while most of people believe that the
infinity comes from the fundamental topological structure of QFT, which can
not be cured without modifying its topological structure. In string theory,
one extends a point particle to a small piece of a string. This simple
modification dramatically changes the topological structure of the theory. The
new "Feynman diagram" now is a smooth world-sheet instead of world-lines with
singular intersection points\textsf{.}

Let us briefly look at the high energy behavior in QFT by a simple power
counting. The tree amplitude of four spin-$J$ particles behaves as
$A_{tree}^{\left(  J\right)  }\sim E^{-2(1-J)}$, so that the one-loop
amplitude behaves as%
\begin{equation}
A_{1-loop}^{\left(  J\right)  }\sim\int d^{4}p\frac{\left(  A_{tree}%
^{(J)}\right)  ^{2}}{\left(  p^{2}\right)  ^{2}}\sim\int d^{4}E\frac
{E^{-4(1-J)}}{E^{4}},
\end{equation}
which is finite for a scalar particle ($J=0$) and renormalizable for a vector
particle ($J=1$), but is nonrenormalizable for the particle with $J\geq2$,
including graviton ($J=2$). However, there is a loophole to bypass this
argument. If we sum over all amplitudes with the different spins, the final
amplitude will be%
\begin{equation}
A=\sum_{J}A_{tree}^{\left(  J\right)  }\sim\sum_{J}a_{J}E^{-2(1-J)},
\end{equation}
which could be exponentially fall-off, i.e. $A\sim e^{-E}$, so that the all
loop amplitudes would be finite, if the following two conditions are both satisfied:

\begin{enumerate}
\item there are infinite higher spin $J$ particles

\item the coefficients $a_{J}$'s\ should be precisely related to each other.
\end{enumerate}

In string theory, the amplitude of four tachyons scattered, i.e. Veneziano
amplitude, can be easily calculated as%
\begin{equation}
\mathcal{T}=B\left(  -\frac{s}{2}-1,-\frac{t}{2}-1\right)  ,
\end{equation}
where $s$, $t$ are the Mandelstam variables. In the high energy and fixed
angle limit (hard scattering where $s\rightarrow\infty$, $t\rightarrow\infty$
and $s/t$ fixed), it is easy to verify that the Veneziano amplitude behaves as
exponentially fall-off. This property applies to all four-point scattering
amplitudes in string theory. Thus string theory is a finite theory! To explain
the reason why string theory is finite, we borrow the previous argument of QFT
and make our conjecture that string theory satisfies the above two conditions
and is a loophole.

The first condition is trivially satisfied in string theory because a string
has infinite oscillation modes which correspond to infinite higher spin
particles. The second condition is highly nontrivial and will lead to the
conjecture of the symmetries among the different string scattering amplitudes.
We conjecture that these symmetries are the key point to "fine-tune" the
string theory to be finite. These symmetries are usually complicated and not
apparent so that we call them hidden symmetries. However, we expect that the
hidden symmetries will reduce to some simpler relations in certain limits. In
fact, Gross has conjectured that the string scattering amplitudes are linearly
related each other in the high energy, fixed scattering angle limit \cite{GM,
Gross, GrossManes}. Using the methods of Ward identities of zero norm states
\cite{ZNS1, ZNS3, ZNS2}, Virasoro algebra and direct calculation of scattering
amplitudes, we are able to prove the Gross conjecture and compute the linear
ratios among the different string amplitudes \cite{ChanLee1, ChanLee2, CHL,
CHLTY, PRL, paperB, susy, Closed, HL, Dscatt, Decay}. We also extend our study
to the high energy, small angle limit, i.e. Regge scattering \cite{bosonic,
bosonic2, RRsusy, LYYan, LMY}. In the following sections, we will review our
results in both limits.

\section{Hard Scattering}

In this section we consider the hard scattering, which is the limit of
$E\rightarrow\infty$ with fixed scattered angle $\phi$, or in Mandelstam
variables $s\sim E^{2}\rightarrow\infty$\textsf{ }with\textsf{ }$t/s\sim
\sin^{2}\left(  \phi/2\right)  $\textsf{ }fixed. It was shown \cite{CHLTY,PRL}
that for the 26D open bosonic string the only four-point scattering amplitudes
that will survive in the hard scattering limit at mass level $M^{2}=\left(
N-1\right)  $ are of the form%
\begin{equation}
\mathcal{T}^{(N,2m,q)}=\langle V_{1}V_{2}^{(N,2m,q)}(k)V_{3}V_{4}\rangle,
\end{equation}
where $V_{1}$, $V_{3}$ and $V_{4}$ are set to be tachyonic vertices for
simplicity and the second vertex of the nontrivial higher spin state is%

\begin{equation}
V_{2}^{(N,2m,q)}(k)\sim\left(  \alpha_{-1}^{T}\right)  ^{N-2m-2q}\left(
\alpha_{-1}^{L}\right)  ^{2m}\left(  \alpha_{-2}^{L}\right)  ^{q}|0;k\rangle.
\end{equation}
the polarizations of the second vertex with momentum $k_{2}$ on the scattering
plane were defined to be $e^{P}=\frac{1}{M_{2}}(E_{2},\mathrm{k}_{2}%
,0)=\frac{k_{2}}{M_{2}}$ as the momentum polarization, $e^{L}=\frac{1}{M_{2}%
}(\mathrm{k}_{2},E_{2},0)$ the longitudinal polarization and $e^{T}=(0,0,1)$
the transverse polarization. Note that $e^{P}$ approaches to $e^{L}$ in the
hard scattering limit, and the scattering plane is defined by the spatial
components of $e^{L}$ and $e^{T}$. Polarizations perpendicular to the
scattering plane are ignored because they are kinematically suppressed for
four-point scatterings in the high-energy limit. One can then use three
different methods, i.e. Ward identities of zero norm states, Virasoro algebra
and saddle point approximation to calculate the ratios among the high energy
scattering amplitudes. The final result of the ratios of the amplitude in hard
scattering limit are \cite{CHLTY,PRL}%
\begin{equation}
\dfrac{\mathcal{T}^{(N,2m,q)}}{\mathcal{T}^{(N,0,0)}}=\left(  -\dfrac{1}%
{2M}\right)  ^{q}\left(  \dfrac{1}{2M^{2}}\right)  ^{m}(2m-1)!! \label{ratios}%
\end{equation}
The methods of Ward identities of zero norm states and Virasoro algebra are
algebraic methods, which in principle can be applied to arbitrary higher loop
level. The\ third method, i.e. saddle point method, is due to direct
calculation of the string scattering amplitudes. To show the linear relations
in the high energy limit, in most of cases, we only calculate the tree level
scattering amplitudes. The direct calculation at higher loop levels is quite
complicated, a simple case at 1-loop level has been showed in \cite{0411212}.

To see the meaning of the ratios in Eq.(\ref{ratios}), let's consider a simple
analogy from practical physics. The ratios of the nucleon-nucleon scattering
processes%
\begin{subequations}%
\begin{align}
(a)\text{ \ }p+p  &  \rightarrow d+\pi^{+},\\
(b)\text{ \ }p+n  &  \rightarrow d+\pi^{0},\\
(c)\text{ \ }n+n  &  \rightarrow d+\pi^{-}%
\end{align}%
\end{subequations}%
can be calculated to be%
\begin{equation}
T_{a}:T_{b}:T_{c}=1:\frac{1}{\sqrt{2}}:1
\end{equation}
from $SU(2)$ isospin symmetry. Similarly, as we will see in the rest of the
paper, the ratios in Eq.(\ref{ratios}) can be extracted from Kummer function.
The key is to study high energy string scatterings in the Regge limit.

\section{Regge Scattering}

In this section we consider the Regge limit, which is the limit of
$E\rightarrow\infty$ with small scattered angle $\phi$, or in Mandelstam
variables $s\sim E^{2}\rightarrow\infty$\textsf{ }with\textsf{ }$t\sim
E^{2}\sin^{2}\left(  \phi/2\right)  $\textsf{ }fixed. Now all four-point
scattering amplitudes survive in the Regge limit at mass level $M^{2}=\left(
N-1\right)  $ with the vertices%
\begin{equation}
V_{2}^{(N,k_{n},q_{m})}(k)\sim\prod_{n>0}(\alpha_{-n}^{T})^{k_{n}}\prod
_{m>0}(\alpha_{-m}^{L})^{q_{m}}|0\rangle, \label{vertex2}%
\end{equation}
where the powers $k_{n}$'s\ and $q_{m}$'s\ satisfy the constraint,

\begin{equation}
\sum_{n,m}nk_{n}+mq_{m}=N.
\end{equation}
The four-point string scattering amplitudes of the above vertex (\ref{vertex2}%
) in the Regge limit can be calculated to be \cite{bosonic}%
\begin{align}
\mathcal{T}^{(N,k_{n},q_{m})}  &  =\left(  -\frac{i}{M_{2}}\right)  ^{q_{1}%
}U\left(  -q_{1},\frac{t}{2}+2-q_{1},\frac{t+M^{2}+2}{2}\right) \nonumber\\
&  \cdot B\left(  -1-\frac{s}{2},-1-\frac{t}{2}\right)  \cdot\prod
_{n=1}\left[  i\sqrt{-t}(n-1)!\right]  ^{k_{n}}\nonumber\\
&  \cdot\prod_{m=2}\left[  i\left(  t+M^{2}+2\right)  (m-1)!\left(  -\frac
{1}{2M_{2}}\right)  \right]  ^{q_{m}}, \label{amplitude in RR}%
\end{align}
which is power-law behaved in the high energy limit as expected. The function
$U(a,c,x)$ in Eq.(\ref{amplitude in RR}) is the Kummer function of the second
kind and is defined to be%
\begin{equation}
U(a,c,x)=\frac{\pi}{\sin\pi c}\left[  \frac{M(a,c,x)}{(a-c)!(c-1)!}%
-\frac{x^{1-c}M(a+1-c,2-c,x)}{(a-1)!(1-c)!}\right]  ,
\end{equation}
where%
\begin{equation}
M(a,c,x)=\sum_{j=0}^{\infty}\frac{(a)_{j}}{(c)_{j}}\frac{x^{j}}{j!},
\end{equation}
is the Kummer function of the first kind. $U(a,c,x)$ and $M(a,c,x)$ are the
two solutions of the Kummer equation%
\begin{equation}
xy^{^{\prime\prime}}(x)+(c-x)y^{\prime}(x)-ay(x)=0. \label{Kummer eq}%
\end{equation}
At this point, it is crucial to note that, in our case of
Eq.(\ref{amplitude in RR}), $c=\frac{t}{2}+2-2m$ and is not a constant as in
the usual definition, so the function $U\left(  -q_{1},\frac{t}{2}%
+2-q_{1},\frac{t+M^{2}+2}{2}\right)  $ in Eq.(\ref{9}) is actually\textit{
not} a solution of the Kummer equation (\ref{Kummer eq}). This makes the
function $U\left(  -q_{1},\frac{t}{2}+2-q_{1},\frac{t+M^{2}+2}{2}\right)  $
has some special properties which will lead to our expected result.

It is important to note that there is no linear relation among high energy
string scattering amplitudes of different string states for each fixed mass
level in the Regge limit as can be seen from Eq.(\ref{amplitude in RR}). This
is very different from the result in the hard scattering limit in
Eq.(\ref{ratios}). In other words, the ratios $\mathcal{T}^{(n,2m,q)}%
/\mathcal{T}^{(n,0,0)}$ are $t$-dependent functions. However, we believe that
the hidden symmetry which we conjectured should be encoded in general string
scattering amplitudes. Therefore the hidden symmetry should also hide in the
scattering amplitudes of the Regge limit in some way. To see the hidden
symmetry in the Regge limit, we take an additional limit\textsf{ }$\left\vert
t\right\vert \rightarrow\infty$ in the scattering amplitudes
(\ref{amplitude in RR}). Using the identity of our "special Kummer function",%
\begin{align}
&  U\left(  -2m,\frac{t}{2}+2-2m,\frac{t}{2}\right) \nonumber\\
&  =0\cdot t^{2m}+0\cdot t^{2m-1}+\cdots\cdots+0\cdot t^{m+1}+\frac
{(2m)!}{\left(  -4\right)  ^{m}m!}t^{m}+O\left(  t^{m-1}\right)
\label{Kummer identity}%
\end{align}
We found that the ratios in Eq.(\ref{ratios}) could be exactly reproduced
among the leading terms of the scattering amplitudes (\ref{amplitude in RR})
under the additional limit.

It can be shown that the identity (\ref{Kummer identity}) is satisfied order
by order for arbitrary value $m$ by computer, but it is highly non-trivial to
prove it analytically. With the help of the mathematicians, we finally
completed the proof two years after we conjectured \cite{LYYan}.

\section{Compactified Space}

It is interesting to study that strings scattered in a compactified space. We
calculated string scattering amplitudes in some compactified spaces for both
hard scattering and Regge limits \cite{Compact, HLY}. Besides the two well
known high energy behaviors of string scattering amplitudes, e.g.
exponentially fall-off in hard scattering and power-law fall-off in Regge
limit, we found two new behaviors under certain conditions:

\begin{enumerate}
\item Power-law fall-off in hard scattering.

\item Exponentially fall-off in Regge limit.
\end{enumerate}

The two new high energy behaviors are completely not expected for the first
sight. But when we carefully study the detail of the kinematic variables, by
comparing the apparent scattering angle $\phi$ in the compactified space and
the effective scattering angle $\phi^{\prime}$ in the full space, we found
that the high energy behaviors are actually determined by the effective
scattering angle $\phi^{\prime}$. We summary our results in the following table:

\begin{center}%
\begin{tabular}
[c]{|c|c|c|c|}\hline
$\phi$ & $\left(  s,t,N,K\right)  $ & $E\rightarrow\infty$ & $\phi^{\prime}%
$\\\hline
\multicolumn{1}{|l|}{$\phi$ \textsf{fixed}} & $\left(  s,t\right)  \gg\left(
N,K\right)  $ & $e^{-E}$ & \multicolumn{1}{|l|}{$\phi^{\prime}$ \textsf{fixed}%
}\\\hline
\multicolumn{1}{|l|}{$\phi$ \textsf{fixed}} & $\left(  s,t,K\right)  \gg N$ &
$E^{-c}$ & \multicolumn{1}{|l|}{$\phi^{\prime}\sim{\small 0}$}\\\hline
\multicolumn{1}{|l|}{$\phi\sim{\small 0}$} & $s\gg\left(  t,N,K\right)  $ &
$E^{-c}$ & \multicolumn{1}{|l|}{$\phi^{\prime}\sim{\small 0}$}\\\hline
\multicolumn{1}{|l|}{$\phi\sim{\small 0}$} & $\left(  s,K\right)  \gg\left(
t,N\right)  $ & $e^{-E}$ & \multicolumn{1}{|l|}{$\phi^{\prime}$ \textsf{fixed}%
}\\\hline
\end{tabular}

\end{center}

If string theory showed to be correct and thus the extra dimensions exist,
then our 3+1 space-time is compactified from a higher dimensional space-time.
The scattering angles measured in our usual scattering experiments should be
the apparent scattering angles. Therefore, under certain conditions as we
showed before, the unexpected high energy behavior could be observed.

\section{Conclusion}

As a theory including quantum gravity, string theory has been showed to be
finite to all loop level. But there are many higher spin modes in string
theory and the scattering amplitudes of each of them are turn out to be
infinite. So there must be a mysterious cancellation in string theory. We
believe that the cancellation is due to the "hidden" symmetries in string
theory. To explore the complicated hidden symmetries, we study the string
scattering amplitudes in different high energy limits. In the hard scattering
limit, we proved that the scattering amplitudes are linear related each other.
We calculated the ratios by three different methods. In the Regge limit, we
calculated the string scattering amplitudes, which is proportional to a
"Kummer" function. By a nontrivial identity of the "Kummer" function, we are
able to show that the string scattering amplitudes in the Regge limit
reproduces to all the ratios in hard scattering limit when we take
$t\rightarrow\infty$.

Our study is just the first step to explore the hidden symmetries in string
theory. To understand the full symmetries, we need to study the string
scattering amplitudes in general energy, but which is very complicated to deal with

\begin{description}
\item[Acknowledgement] 
\end{description}

Yi would like to thank the organizers of 11th workshop on QCD for inviting him
to present this work. This work is supported by the National Science Council
and National Center for Theoretical Science, Taiwan.

\begin{center}

\end{center}

\end{document}